# On the Performance Potential of Speculative Execution Based on Branch and Value Prediction

Pece Mitrevski and Marjan Gušev

**Abstract:** Fluid Stochastic Petri Nets are used to capture the dynamic behavior of an ILP processor, and discrete-event simulation is applied to assess the performance potential of predictions and speculative execution in boosting the performance of ILP processors that fetch, issue, execute and commit a large number of instructions per cycle.

## 1 Introduction

Current microprocessor architectures assume sequential programs as an input and a parallel execution model. Their efficiency is highly dependent on the Instruction-Level Parallelism (ILP) the programs exhibit, as a measure of the potential number of instructions that can be executed simultaneously. Basically, there are two fundamental approaches to executing multiple instructions in parallel: the superscalar processor and the VLIW (Very Long Instruction Word) processor. The superscalar approach extracts the parallelism from a sequential program at run time (dynamically), by employing sophisticated hardware mechanisms. On the other hand, the performance of VLIW architectures is dependent on the capability of the compiler to detect and exploit instruction-level parallelism during instruction scheduling using advanced (static) techniques.

Regardless of the approach used, instructions cannot always be eligible for parallel execution due to three classes of constraints: control dependences, true data dependences and name (false) dependences [5]. Control dependences appear when the execution of instructions depends on the outcome of a branch (either conditional or unconditional). The resolution of the outcome involves deciding whether the branch is taken or not (in the case of conditional branch) and computing its target address. Control dependent instructions cannot even begin their execution because the branch outcome should determine the next sequence of instructions to be executed. True data dependences occur when an instruction consumes a value that is generated by a preceding instruction, and therefore these instructions cannot be executed in parallel. Name dependences occur when instructions use the same register or memory location (name), but there is no flow of data between them as in true data dependences.

Many ILP processors speculatively execute control dependent instructions before resolving the branch outcome. They rely upon branch prediction in order to tolerate the effect of control dependences. A branch predictor uses the current fetch address to predict whether a branch will be fetched in the current cycle, whether that branch will be taken or not, and what the target address of the branch is. The predictor uses this information to decide where to fetch from in the next cycle. Since the branch execution penalty is only seen if the branch was mispredicted, a highly accurate branch predictor is a very important mechanism for reducing the branch penalty in a high performance ILP processor.

Given that a majority of static instructions exhibit very little variations in values that they produce/consume during the course of a programs execution, data dependences can be eliminated at run-time by predicting the outcome values of instructions (value prediction) and by executing the true data dependent instructions. In general, the outcome value of an instruction can be assigned to registers, memory locations, condition codes, etc. The execution is speculative, as it is not assured that consumer-instructions were fed with correct input values. Since the correctness of the execution must be maintained, speculatively executed instructions retire only if the predictions they rely upon were proven correct - otherwise, they are discarded and reissued with the correct values.

## 2 Motivation and Related Work

The aim of a plethora of works has been studying the limits of ILP, i.e. the influence that predictions and speculative execution have on ILP processor performance under different scenarios (microarchitectural features): instruction fetch bandwidth, prediction accuracy, available resources, instruction window size, issue width, etc. More or less, they all rely upon the use of microarchitectural simulators [1,4]. An alternative is the analytical modeling approach - a set of formulas or equations describe the system, and manipulating or solving the equations leads to results that describe the system behavior. In simpler cases, equations can be solved to get a closed-form answer, but more often, a numerical solution needs to be carried out. Analytical models are generally more of an abstraction of the system than the microarchitecture models used in simulators. Nevertheless, the models published so far do not even distantly capture the dynamic behavior of an ILP processor with speculative execution based on predictions. Only a few deterministic models are known [3, 6, 8] - they deal with average parameter values (or a parameter takes only one value), there is no randomness and the result is based on known functions of inputs with no dependence on chance. Some authors [6] point out that these models provide some insight by isolating important parameters, but they are still too simple to capture the behavior of a real system.

Opposed to this common trait, a stochastic model of the dynamic behavior of an ILP processor that aims to minimize the impact of control and true-data dependences and employs speculative execution based on branch and value prediction, has been introduced for the first time in [7]. In view of the fact that in a machine with multiple execution units capable to execute large number of instructions in parallel the service and storage requirements of each instruction are small compared to the total volume of the instruction stream, individual instructions are regarded as atoms of a fluid and large buffer levels are approximated by continuous fluid levels. As a result, state-space complexity is decreased. The dynamic behavior model is built using Fluid Stochastic Petri Nets (FSPN) [9] and the stochastic process underlying the Petri Net is described by a system of first-order hyperbolic partial differential equations with appropriately chosen initial and boundary conditions.

In this paper, we present performance evaluation results obtained using discrete-event simulation [2] of a slightly modified FSPN model, in order to better understand branch and value prediction techniques and their performance potential with varying machine width.

## 3   The FSPN Model

FSPNs contain two types of places: discrete places containing a non-negative integer number of tokens, and continuous places containing fluid (non-negative real quantity). Transition firings are determined by both discrete and continuous places, and fluid flow is permitted either with deterministic fluid rates through the enabled timed transitions, or in the form of fluid jumps (transportation of fluid in zero time) through enabled immediate transitions in the Petri Net. From the FSPN point of view, despite the microarchitectural complexity, fairly simple concept lies beneath the dynamic behavior model: instructions flow and pass through separate pipeline stages connected by buffers. Control dependences stall the inflow of useful instructions (fluid) into the pipeline, whereas true data dependences decrease the aperture of the pipeline and the outflow rate. The buffer levels always vary and affect both the inflow and outflow rates. The speculative execution based on branch prediction tends to eliminate stalls in the inflow, while the speculative execution based on value prediction helps keeping the outflow rate as high as possible.

Based on the assumption that the pipeline is organized in, more or less, four stages - Fetch, Decode/Issue, Execute and Commit, fluid places $P_{IC}$, $P_{IB}$, $P_{RS/LSQ}$, $P_{ROB}$, $P_{RR}$, $P_{EX}$ and $P_{REG}$, depicted by means of two concentric circles (Fig. 1), represent buffers between pipeline stages - instruction cache, instruction buffer, reservation stations & load/store queue, reorder buffer, rename registers, instructions that have completed execution and architectural registers. The fluid place $P_{TIME}$ has the function of an hourglass: it is constantly filled at rate 1 up to the level 1 and then flushed out, which corresponds to the machine clock cycle. Fluid arcs are drawn as double arrows to suggest a pipe. Flow rates are piecewise constant, i.e. take different values at the beginning of each cycle and are limited by the fetch/issue width of the machine ($W$). Rates depend on fluid levels and change when $T_{CLOCK}$ fires and the fluid in $P_{TIME}$ is flushed out. The flushout arc is drawn as thick single arrow. A high-bandwidth instruction fetch mechanism fetches up to $W$ instructions per cycle with rate $r_{FETCH}$ and places them in the instruction buffer. In the case of a branch misprediction, the fetch unit is effectively stalled and no useful instructions are added to the buffer. Instruction cache misses are ignored. Instruction issue tries to send $W$ instructions to the appropriate reservation stations or the load/store queue on every clock cycle. The actual issue rate is $r_{ISSUE}$. Rename registers are allocated to hold the results of the instructions and reorder buffer entries are allocated to ensure in-order completion. Among the instructions that initiate execution in the same cycle, speculatively executed consumer-instructions are forced to retain their reservation stations. Up to $W$ instructions are in execution at the same time. With the assumption that functional units are always available and out-of-order execution is allowed, the instructions initiate and complete execution with rate $r_{INITIATE} = r_{COMPLETE}$. During the execute stage, the instructions first check to see if their source operands are available (predicted or computed). For simplicity, the execution latency of each instruction is a single cycle. Instructions execute and forward their own results back to subsequent instructions that might be waiting for them (no result forwarding delay). Every reference to memory is present in the first-level cache - the effect of the memory hierarchy is eliminated. The instructions that have completed execution are ready to move to the last stage. Up to $W$ instructions may commit per cycle. The results in the rename registers are written into the register file and the rename registers and reorder buffer entries freed with rate $r_{COMMIT}$.

Initially, tokens occur in places $P_{FETCH}$ and $P_{INITIATE}$, while the fluid place $P_{IC}$ is filled with fluid with volume $V$, equivalent to the total number of useful instructions (program volume). Firing of exponential transition $T_{BRANCH}$ corresponds to a branch instruction occurrence. The parameter $\lambda$ changes at the beginning of each clock cycle and depends on the fetch rate: $\lambda = \lambda_i * r_{FETCH}/W$, where $\lambda_i$ is its upper limit at maximum fetch rate ($r_{FETCH} = W$). The branch is correctly predicted with probability $1 - p_{BMIS}$, or mispredicted with probability $p_{BMIS}$. These probabilities are included in the FSPN model as weights assigned to immediate transitions $T_{BPC}$ and $T_{BPMIS}$, respectively. This is known as synthetic branch prediction. Branch mispredictions stall the fluid inflow for as many cycles as necessary to resolve the branch ($C_{BR}$ tokens in place $P_{BMIS}$). Usually, a branch is not resolved until its execution stage ($C_{BR} = 3$). With several consecutive firings of $T_{CLOCK}$, these tokens are consumed one at a time and moved to $P_{RESOLVED}$. As soon as the branch is resolved, transition $T_{CONTINUE}$ fires, a token appears in place $P_{FETCH}$ and the inflow resumes. Similarly, firing of exponential transition $T_{CONSUMER}$ corresponds to the occurrence of a consumer-instruction among the instructions that initiated execution. The parameter $\mu$ changes at the beginning of each clock cycle and depends on the initiation rate: $\mu = \mu_i * r_{INITIATE}/W$, where $\mu_i$ is its upper limit when maximum possible number of instructions simultaneously initiate execution ($r_{INITIATE} = W$).

The consumed value was correctly predicted with probability $1 - p_{VMIS}$,

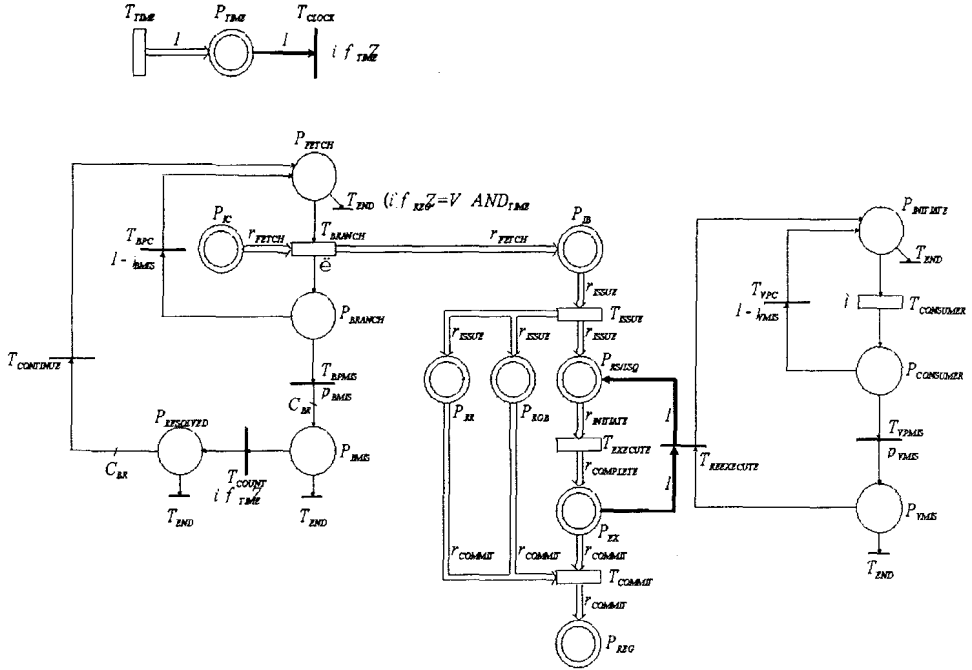

Fig. 1. A fluid stochastic petri net model.

or mispredicted with probability $p_{VMIS}$. These probabilities are included in the FSPN model as weights assigned to immediate transitions $T_{VPC}$ and $T_{VPMIS}$, respectively. Whenever a misprediction occurs (token in place $P_{VMIS}$), the consumer-instruction has to be rescheduled for execution. The firing of immediate transition $T_{REEXECUTE}$ causes transportation of fluid in zero time. Fluid jumps have deterministic height of 1 (one instruction). Jumps that would go beyond the boundaries cannot be carried out. The arcs connecting fluid places and immediate transitions are drawn as thick single arrows. The fluid flow terminates at the end of the cycle when all the fluid places except $P_{REG}$ are empty and $T_{END}$ fires.

## 4   Performance Evaluation Results

We investigate branch and value prediction efficiency with varying machine width (Figs. 2 and 3). The results have been obtained using discrete-event simulation, specifically implemented for this model and not for a general FSPN. The types of events that need to be scheduled in the event queue are either transition firings or the hitting of a threshold dependent on fluid levels. We have used a Unif[0,1] pseudo-random number generator to generate samples from the respective cumulative distribution functions and determine firing times of timed transitions via inversion of the *cdf* ("Golden Rule for Sampling"). In the case of branch prediction (Fig. 2), the speedup is computed by dividing the IPC achieved in a machine over the IPC achieved in a scalar counterpart ($W = 1, \mu_i = 0$). Program volume is set to $V = 10^6$ instructions, while fluid places capacity is $2W$. The speedup due to branch prediction is obviously higher in wider machines. With perfect branch prediction, the speedup unconditionally increases with the machine width. For a given width, the speedup is higher when there are a smaller number of consuming instructions (low $\mu_i/W$). With realistic branch prediction, there is a threshold effect on the machine width: below the threshold the speedup increases with the machine width, whereas above the threshold the speedup is close to a limit - machine width is by far larger than the average number of instructions provided by the fetch unit. The threshold decreases with increasing the number of mispredicted branches. This is in agreement with the results reported in [6] where a threshold effect on the instruction fetch rate was exposed.

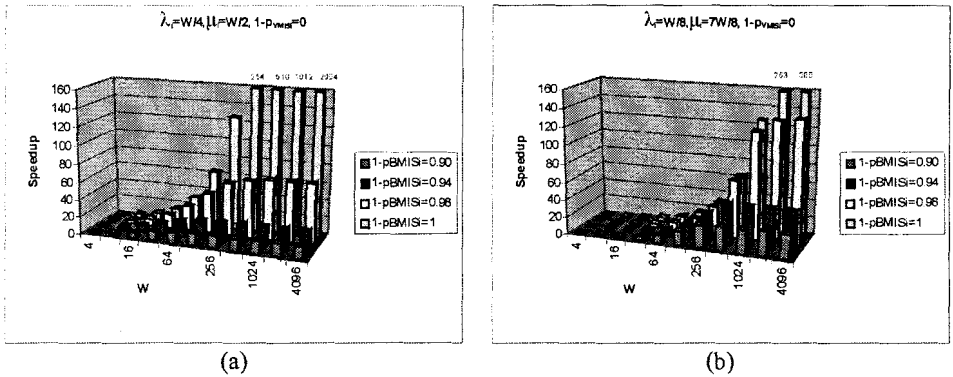

Fig. 2. Speedup achieved by branch prediction with varying machine width.

The maximum additional speedup that value prediction can provide is computed by dividing the IPC achieved with perfect value prediction over the IPC achieved without value prediction (Fig. 3). With perfect branch prediction, some true data dependences can always be eliminated, regardless of the machine width. Actually, the maximum additional speedup is predetermined by the ratio $W/(W - \mu_i)$. However, with realistic branch prediction, the additional speedup diminishes when the machine width is above a threshold value. It happens earlier when there are a smaller number of consuming instructions and/or a larger number of mispredicted branches. In either case, the number of independent instructions examined for simultaneous execution is sufficiently higher than the number of fetched instructions that enter the instruction window. Branch prediction becomes more important with wider processors.

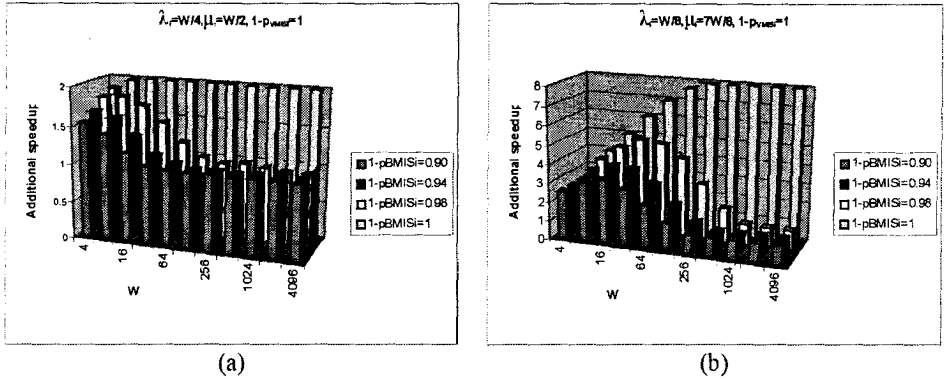

Fig. 3. Maximum additional speedup achieved by perfect value prediction with varying machine width.

## 5 Conclusion

The main conclusions that can be drawn from this study are the following:

- The benefits of branch and value prediction are higher when control and true data dependences have a much higher influence on the total execution time of a program;
- Value prediction is an effective approach that might enable higher levels of parallelism without the need to increase machine width;
- There is a correlation between the value prediction efficiency and the branch prediction efficiency;
- The wider the machine, the more significant performance limitation the branch mispredictions become.